
\magnification=\magstep1
\hsize 32 pc
\vsize 42 pc
\baselineskip = 24 true pt
\centerline {\bf Exact Solutions to the Generalised Lienard Equation}
\vskip .5 true cm
\centerline {\bf B. Dey$^*$}
\centerline {Department of Physics, University of Poona,}
\centerline {Ganeshkhind, Pune 411 007, India}
\vskip .3 true cm
\centerline {\bf Avinash Khare$^{**}$}
\centerline {Institute of Physics, Sachivalaya Marg,}
\centerline { Bhubaneswar - 751 005, India}
\vskip .3 true cm
\centerline {\bf C. Nagaraja Kumar}
\centerline {School of Physics, University of Hyderabad,}
\centerline { Hyderabad - 500 046, India}
\vfill

\noindent {\bf Abstract}

Many new solitary wave solutions of the recently
studied Lienard equation are obtained by mapping it to the field equation
of the $\phi^6-$field theory. Further, it is shown that the exact
solutions of the Lienard equation are also the exact solutions of the
various perturbed soliton equations. Besides, we also consider a one parameter
family of generalised Lienard equations and obtain exact solitary wave
solutions of these equations and show that these are also the
exact solutions of the various other generalised nonlinear equations.
\vskip .5 true cm

* \ e-mail address : bdey@physics.unipune.ernet.in

**  \ e-mail address : khare@iopb.ernet.in

\eject
\noindent {\bf I. Introduction}

Recently Kong [1] has obtained an exact solitary wave
solution of the Lienard equation. The study of the Lienerd equation
is very useful because finding solitary wave solutions of some
important equations such as RR equation [2], the A-equation [3] and the GI
equation [4] which are proposed as integrable conditions for the
nonlinear Schr\"odinger equation and also host of other equations [1]
can be reduced to finding the solutions of the Lienard equation.
Moreover, as we will show in this letter, exact solutions of various
perturbed soliton equations like cubic nonlinear
Schr\"odinger equation, KdV equation as well as the recently
considered Wadati-Segur-Ablowitz equation [5] can also be obtained by
reducing these perturbed equations to the Lienard form.

With these applications in mind,
in this letter we report various new topological and
periodic solitary wave solutions of the Lienard equation. These
new solutions are obtained by identifying the Lienard
equation with the equation of motion of the well known nonlinear
$\phi^6$ field theory in (1+1) dimensions. The analysis of
the $\phi^6$ field theory has been done in great
detail in [6] and this identification results in several new exact
solutions of the Lienard equation. In particular, it turns out that the most
general solution of the Lienard equation can be expressed in terms of
Weirstras\'s function. Besides, we also obtain several exact
solutions of a one parameter family of
the generalised Lienard equations and show that the
exact solutions of the other generalised equations like the generalised
RR-equation, the generalised A-equation, the generalised GI-equation
and the host of
other generalised nonlinear equations can be expressed in terms of the
exact solutions of this equation.
\vskip .4 true cm
\noindent {\bf II. New solitary wave solutions of the Lienard equation}

The Lienard equation is given by [1]
$$\phi^{''}(\zeta)+l\phi(\zeta) + m\phi^3(\zeta)+ n\phi^5(\zeta) =0
\eqno{(1)}$$
where l,m and n are constant coefficients. Kong has obtained a
nontopological solitary wave solution of this equation of the form
$$\phi(\zeta) =  {1\over {[A + B \cosh {D(\zeta +\zeta_0)}]}^{1/2}}\eqno{(2)}$$
in case either $l <0, n >0$ or $l,n <0, m>0$ and $nl/m^2 < 3/16$.

Let us first show that the Lienard equation can be identified with the
field equations of the $\phi^6$
nonlinear field theory. which in (1+1) dimensions
is used as a model for structural phase transition [6].
The onsite potential corresponding to this model is given by
$$V(\phi_i) = B\phi_i^2 + A\phi_i^4 + C\phi_i^6\eqno{(3)}$$
In the continuum limit, the corresponding equation of motion is given by [6]
$$m{\partial^2\phi\over\partial t^2}-m C^2_0
{\partial^2\phi\over\partial x^2} + 2B\phi + 4A\phi^3 + 6C\phi^5
= 0\eqno{(4)}$$
In the travelling wave frame $\zeta = x-vt, \ \phi(x,t) =
\phi(\zeta)$ and this equation reduces to
$${d^2\phi\over dS^2}-\phi- {2A\over B}\phi^3 - {3C\over B}\phi^5 =
0\eqno{(5)}$$
where $S = \zeta/\zeta_0$ and $\zeta^2_0 = m (C^2_0-v^2)/2B$.
 Thus this $\phi^6$-field equation is nothing but the Lienard
eq. (1) considered by Kong [1], for the choice of the
coefficients $l = - 1, m = - 2A/B$ and $n = - 3C/B$.
Accordingly, all the solutions of eq. (5) will also be the solution
of the Lienard equation (1) for the suitable values of the coefficients
$l, m$ and $n$.

The onsite-potential $V(\phi_i)$ as given by eq. (3) whose average value
essentially
corresponds to the static free energy of the system exhibits a first
order phase transition [6] if
$ B,C > 0, A < 0 \ and \ 0 < a < 3/2$
where $a={9BC\over 2\mid A\mid^2}$. Under these conditions the
potential has three minima. On the other hand if $B <0, C >0$ then the
potential
has two degenerate minima irrespective of the sign of A. Finally, in case $C
<0$
, then the Lienard equation can be identified with the equation of motion of a
classical particle (of unit mass) moving in an anharmonic potential [7].

Various solutions of eq. (5) have been presented in [6] and [7]. In particular,
it has been shown that the exact solution of eq. (5) can be expressed
in terms of the Weirstras\'s function. In view of the above identification, we
thus see that the most general solution of the Lienard eq. (1) can be written
in terms of Weirstras\'s function. On borrowing the results of [6,7] some
simple special solutions of the Lienard equation (1) are

\item {(i)} the topological solitary wave (kink or domain wall) solutions
which are given by
$$\phi(\zeta) = \pm {A\tanh (Bx) \over {[1-C\tanh^2(Bx)]^{1/2}}}\eqno{(6)}$$
where A = $[-2l(1-3C)(1-C)/m(2-3C)]^{1/2}$, B = $[l/(2-3C)]^{1/2}$ and
$ln/m^2 = -3C(2-3C)/[2(1-3C)]^2$. Here C
is an arbitrary constant with $C< 1$ so that the solution is nonsingular
while x = $(\zeta+\zeta_0)$. As x
goes from $-\infty$ to
$+\infty$, $\phi(\zeta)$ goes from one absolute minima of the
potential to the other. Notice that  (i) if $2/3 < F <1$ then $n,l < 0, m >0$
 (ii) if $1/3 < C < 2/3$ then $n <0, l,m >0 $ (iii) if $0 < C < 1/3$ then
$n,m <0 , l >0$ (iv) if $C < 0$ then $n,l > 0 , m < 0$.

\item{(ii)} For $n,l > 0, m <0$ and $nl/m^2 = 1/4$ there is a special
solution given by
$$\phi(\zeta) = \pm {(-2l/m)^{1/2}x \over {[{(3/l)}+x^2]^{1/2}}}.\eqno{(7)}$$
Apart from these, there are few other exact solutions of the Lienard eq. (1)
which we shall mention below in Sec.4 when we discuss the various solutions
of the generalised Lienard equations.
As a corollary, we can say that the RR-equation, the A-equation and the
GI-equation and the host of other equations [1] which can be reduced
to the Lienard eq. (1) will also have the new exact solitary
wave solutions as given above as well as in Sec.4
\vskip .4 true cm
\noindent {\bf III. Applications: Perturbed Soliton Equations}

Perturbed soliton equations represent those which differ slightly
from the standard soliton equations and represent physically the
more realistic experimental situations, especially the effect of various
forms of dissipation, dispersion etc. which are treated as
perturbation. Various perturbative methods have been developed to
study these perturbed soliton equations [8]. In this section we
shall show that some of the well known perturbed soliton
equations can be reduced to the Lienard form (1) by
using suitable ansatz. Thus the exact solutions of the Lienard
eq. (1) also represent the exact solutions of
these perturbed equations.

{\bf \item {(i)} The Perturbed Cubic Nonlinear Schr\"odinger Equation:}

This equation is represented by [8]
$$i\phi_t + \phi_{xx} + 2\mid \phi\mid^2 \phi = i \in R(\phi)\eqno{(8)}$$
where $\in << 1$ and $R(\phi)$ is some specified function of the solution
$\phi(x,t)$ representing the perturbation.  For example, $\in R(\phi) =\gamma
\phi$
provides a simple description of dissipative process and $\in R(\phi)
=\gamma \phi_{xx}$ induces diffusion effects on the system. Here we
consider $R(\phi) = \mid \phi\mid^2\phi_x$.
To get the exact solutions of the perturbed eq. (8) we use the
ansatz
$$\phi(x,t) =a(\zeta)exp(i[\psi(\zeta)-wt]) \eqno{(9)}$$
where $\zeta=x-vt$.
Substituting eq. (9) in eq. (8) we get
$$-va'+2a'\psi'+a\psi''-\in a^2a' =0\eqno{(10)}$$
and
$$va\psi'+wa+a"-a\psi'^2+2a^3+\in a^3\psi' =0\eqno{(11)}$$
Following Kong [1], let $\psi'(\zeta) = E+Da^2(\zeta)$ where E and D are
constants.
Substituting this in eq. (10) we get $E = v/2$ and $D = \in/4$. Similarly
substitution of $\psi'(\zeta)$ in eq. (11) gives
$$a" + (w+{v^2\over 4})a+(2+{\in v\over 2})a^3 +{3\over 16}\in^2 a^5
= 0\eqno{(12)}$$
which is nothing but the Lienard eq. (1) for the choice of the
coefficients $l = (w+{v^2\over 4}), m = (2+{\in v\over 2})$ and $n =
{3\over 16}$. Thus, for the appropriate choice of the constant
coefficients, various exact solutions of the perturbed cubic nonlinear
Schr\"odinger eq. (8) are immediately obtained in terms of the exact
solutions of the Lienard eq. (1) as given in Secs. 2 and 4 (below).

{\bf \item {(ii)} Perturbed Modified KdV Equation:}

Similarly consider the perturbed modified
KdV equation
$$\phi_t+\alpha \phi^2\phi_x+\gamma \phi_{xxx} = \in \phi^4 \phi_x\eqno{(13)}$$
where the effect of higher order nonlinear term is considered as a
perturbation to the well known integrable modified KdV equation
$\phi_t+\alpha \phi^2\phi_x+\gamma \phi_{xxx}=0$. Using
the transformation $\zeta= x-vt$, eq. (13) can be written as
$$-v\phi_{\zeta}+\alpha \phi^2\phi_{\zeta} + \gamma \phi_{\zeta\zeta\zeta}
 = \in \phi^4\phi_{\zeta}\eqno{(14)}$$
Integrating w.r.t $\zeta$ we get
$$\gamma \phi_{\zeta\zeta} - v\phi +{\alpha\over 3} \phi^3-{\in\over 5}\phi^5
 = 0\eqno{(15)}$$
which is of the Lienard form (1) with the identification of the
coefficients $l = -{v\over\gamma}, m = {\alpha\over 3\gamma}$ and $n=
-{\in\over 5\gamma}$. Hence various exact solutions of the perturbed
modified KdV eq. (13) are also known in terms of the exact
solutions of the Lienard equation as given in Secs. 2 and 4 (below).

{\bf \item {(iii)} Perturbed Wadati-Segur-Ablowitz (WSA) equation:}

Recently Wadati et al. [5] have introduced the equation
$$iu_x+u_{tt}+2\sigma\mid u^2\mid u - \in u_{xt} = 0\eqno{(16)}$$
to study certain instabilities of the modulated wave trains and
obtained an exact solitary wave solution of this equation. Later Kong
and Zhang [9] obtained another exact solitary wave solution of this
equation. Here, we point out that if the WSA equation is perturbed by
a term of the form $\mid u\mid^2 u_x$, then the resulting perturbed WSA
equation can be reduced to the Lienard form
by using the ansatz for u(x,t) as given in eq. (9) and following the
procedure as used for the perturbed cubic nonlinear
schr\"odinger equation.

Yet another form of the perturbed WSA equation which can be casted in the
Lienard form is given by
$$iu_x+u_{tt}+2\sigma \mid u\mid^2 u - \alpha u_{xt} = \in \mid
u\mid^4 u\eqno{(17)}$$
where $\in << 1$ and the higher order nonlinear term is considered as
a perturbation to the WSA eq. (16). Using an ansatz [10] which is
differtent from that in eq. (9) i.e.
$$u(x,t) = e^{i\eta} \phi(\rho)\eqno{(18)}$$
where $\rho = ax-vt $ and $\eta = Kx-\omega t$ and substituting in
eq. (17) one again obtains a Lienard equation.

We would also like to point out that in the WAS eq. (16) if we
replace the $\mid u\mid^2 u$ term by $i\mid u\mid^2 u_x$ term, then
the resulting equation, which we term as modified WSA equation,
can also be reduced to the Lienard form by using the ansatz for
u(x,t) as in eq. (9).
\vskip .4 true cm
\noindent {\bf IV. The Generalised Lienard Equations and their Solutions}

{\bf \item {(1)} The Generalised Lienard Equations:}

Inspired by the success of the Lienard eq. (1), we consider a one
parameter family of
generalised Lienard equations with p'th order nonlinearity i.e. consider
$$\phi^{''}(\zeta)+l\phi(\zeta) + m\phi^{p+1}(\zeta)+
n\phi^{2p+1}(\zeta) =0 \eqno{(19)}$$
where $l$,m and n are constant coefficients and p =1,2,3.... The Lienard
equation in [1]
corresponds to the p=2 case of these generalised equations. As compared to
the onsite potential for the Lienard equation as given by eq. (3), the
generalised Lienard
equation above corresponds to an onsite potential
$$V(\phi_i) = B\phi^2_i+A\phi^{p+2}_i+C\phi^{2p+2}_i,\eqno{(20)}$$
Some exact solutions of the generalised Lienard eqs. (19) and their
applications have been reported in [11]. We now present four
exact solutions of eqs. (19).

(a) The exact topological
solutions of eq. (19) can be written as [11]
$$\phi^{p}(\zeta) = \phi_0 [ 1\pm \tanh Bx], \eqno{(21)}$$
in case $l,n <0$ and $ln/m^2 = (p+1)/(p+2)^2$. Here $\phi_0 =
{-l(p+2)}/{2m}$, B = p$\sqrt{-l}/2$ while x is as defined in Sec. 2.
Notice that if p is even then $m > o$ while if p is odd then it could be of
either sign. For the special case of p = 2 (Lienard equation) we then obtain
the well known [12] topological solution. Notice that for any even (odd) p,
the topological solution has been obtained at the point where the potential
(eq. (20)) has three (two)
degenerate minima.

(b) Another exact solution of eq. (19) is given by
$$\phi(\zeta) = {1\over {[A + B\cosh Dx]^{1/p}}}, \eqno{(22)}$$
where $A= -m/[l(p+2)], B= [{m^2\over l^2(p+2)^2}- {n\over l(p+1)}]^{1/2}$
and $D = p\sqrt{-l}$. Note that this solution is acceptable if either
$l,n <0, m >0$ and $ln/m^2 < (p+1)/(p+2)^2$
or $l<0, n>0$  and m arbitrary. For p=2, this reduces to the solution obtained
by Kong [1]. For odd p, solution with B replaced by -B is also allowed in
either of the two cases. Further, in both cases m could be either positive
or negative.

(c) Yet another exact solution of eq. (19) is the periodic solution given by
$$\phi(\zeta) = {1\over {[A \pm B\sin Dx]^{1/p}}}, \eqno{(23)}$$
where A and B are as given above (below eq. (22)) while $D = p\sqrt{l}$.
This is an acceptable solution if  $l,n >0, m <0$ and
$nl/m^2 <(p+1)/(p+2)^2$. Note that for odd p, m could have either sign.
For p = 2 it gives us the periodic solution of the Lienard eq. (1). It may
however be noted that as shown in [6,7], the most general periodic solution
to the Lienard eq. (1) is infact the Jacobi elliptic function.

(d) For the special case of $l =0, n >0, m < 0$ there is an exact solution
of eq. (19) given by
$$\phi(\zeta) = {1\over {[A + Bx^2]^{1/p}}}\eqno{(24)}$$
where $A = -n(p+2)/[2m(p+1)]$ and $B = - mp^2/[2(p+2)]$. Note again that for
p odd, m can either be positive or negative.

{\bf \item {(2)} The Generalised RR-equation:}

We consider the following generalised RR-equation
$$\phi_{xt} - \beta_1\phi_{xx} + \phi + iT\beta_2{\mid\phi\mid}^p\phi_x = 0,
\ p=1,2,3...\eqno{(25)}$$
To get the exact solution of this equation we use the ansatz [1] as given by
eq. (9) and obtain
$$\phi_t= [-iva'\psi'-iwa-va']exp[i(\psi(\zeta)-wt)]\eqno{(26)} $$
Similarly, we can obtain $\phi_x,\ \phi_{xt}$ and $\phi_{xx}$ which are also
given in [1]. Substituting these values in eq. (25) we get
$$-(v+\beta_1)[a(\zeta)\psi''(\zeta)+2a'(\zeta)\psi'(\zeta)]-wa'(\zeta)
+T\beta_2 a^p(\zeta)a'(\zeta)=0, \eqno{(27)}$$
and
$$-(v+\beta_1)[a''(\zeta)-a(\zeta){\psi'}^2(\zeta)]+wa(\zeta)\psi'(\zeta)
-T\beta_2 a^{p+1}(\zeta)\psi'(\zeta)+a(\zeta)=0 \eqno{(28)}$$
Let
$$\psi'(\zeta)=E+Da^p(\zeta)\eqno{(29)}$$
where E and D are constants. Substituting this in eq. (27) we get
$$E=-{w\over {2(v+\beta_1)}} \ \ and \ \ D={T\beta_2\over
{(p+2)(v+\beta_1)}}\eqno{(30)}$$
 From eq. (28) we then obtain
$$a''(\zeta)-{[4(v+\beta_1)-w^2]\over {4(v+\beta_1)^2}} a(\zeta)-
{T\beta_2 w\over {2(v+\beta_1)^2}}a^{p+1}(\zeta)
+{(p+1)\beta_2^2\over {(p+2)^2(v+\beta_1)^2}} a^{2p+1}(\zeta)=0\eqno{(31)}$$
This is of the form of the generalised Lienard eq. (19) and thus the
various exact solutions of the generalised RR-equation can be expressed in
terms of the exact solutions of the generalised Lienard eq. 919) as mentioned
above. Note that for p=2 case, the generalised RR eq. (25) reduces to the
well known RR equation [2] (see eq. (21) of [1]).

Proceeding in the same way, one can show that the generalised A-equations [3]
$$i\phi_t=\phi_{xx}-4i\mid\phi\mid^{p-2}\phi^2\bar\phi_x+8\mid\phi\mid^{2p}\phi
\eqno{(32)}$$
as well as the generalised GI-equations [4] as given by
$$i\phi_t+\phi_{xx}+2i\delta\mid\phi\mid^{p-2}\phi^2\bar\phi_x
+\beta\mid\phi\mid^p\phi+2\delta^2\mid\phi\mid^{2p}\phi=0\eqno{(33)}$$
can be reduced to the generalised Lienard eq. (19).

Similarly, there are other generalised equations whose exact solutions can also
be obtained
by reducing them to the generalised Lienard eq. (19). Some of these are [1]
$$(i) \ i\phi_t+\phi_{xx}\pm i(\mid\phi\mid^p\phi)_x=0$$
$$(ii) \ i\phi_t+\phi_{xx}\pm i\alpha(\mid\phi\mid^p\phi)_x\mp \beta\mid
\phi\mid^p\phi=0$$
$$(iii) \ i\phi_t+\phi_{xx}\pm i\delta\mid\phi\mid^p\phi_x=0$$
$$(iv) \ i\phi_t+\phi_{xx}+i\alpha(\phi\bar\phi_x+ \bar\phi\phi_x)
\mid\phi\mid^{p-2}\phi=0$$
$$(v) \ i\phi_t+(\phi_{xx}\pm\phi_{yy})+\beta\mid\phi\mid^p\phi+\delta\mid
\phi\mid^{2p}\phi-i\delta div(\mid\phi\mid^p\phi)=0$$

\vskip .2 true cm
\noindent {\bf V. Conclusion}

In conclusion, we have obtained several new solitary
wave solutions of the Lienard equation recently considered by Kong
[1]. An important observation we have made here is the identification
of the Lienard equation with the $\phi^6$ field theory in (1+1)
dimension which describes structural phase transition in various
physical systems [6]. This identification has enabled us to find
many new solutions of the Lienard eq. (1). As an application, we have
 shown how the exact solutions of
the various perturbed soliton equations can also be obtained by reducing
these equations to Lienard form by using some appropriate ansatz.

Very recently, soliton solutions of the complex Ginzberg-Landau equation
have been obtained[13]. We wish to point out that the eq. (3) of their paper
[13] is immediately obtained from Lienard eq. (1) by multiplying it by $\phi$,
 integrating it and choosing the constant of integration to be zero. Thus the
four exact solutions of the Lienard eq. (1) as given by eqs. (21) to (24) (with
p = 2) are also the exact solutions of the complex Ginzburg-Landau eq. (3)
of ref.[13].

Finally, in view of the
interest to study the generalised soliton equations [10],
we have considered a one parameter family of the generalised Lienard eqs. (19)
and obtained its
several exact solutions. As an application,
we have also studied the generalised RR equation,
A-equation, GI-equation and host of other generalised equations and
shown that all of them can be reduced to the generalised Lienard
eqs. (19) so that the exact solutions of these generalised equations
can also be immediately obtained.
\vskip .5 true cm
\eject
\noindent {\bf References}
\item {[1]} D. Kong, Phys. Lett.  {\bf A196} (1995) 301.
\item {[2]} A.A. Rangwala and J.A. Rao, J. Math. Phys. {\bf 31} (1990) 1126.
\item {[3]} M.J. Ablowitz, A. Ramani and H. Segur, J. Math. Phys. {\bf 21}
(1980) 1006.
\item {[4]} V.S. Gerdjikov and I. Ivanov, Bulg. J. Phys. {\bf 10} (1983) 130.
\item {[5]} M. Wadati, H. Segur and M.J. Ablowitz, J. Phys. Soc. Japan
{\bf 61} (1992) 1187.
\item {[6]} S.N. Behera and A. Khare, Pramana {\bf 15} (1980) 245.
\item {[7]} S.N. Behera and A. Khare, Lett. Math. Phys. {\bf 4} (1980)
153; A. Khare and S.N. Behera, Pramana {\bf 14} (1980) 327.
\item {[8]} G.L. Lamb, Jr., Elements of Soliton Theory,
(Wiley-Interscience 1980) p.259.
\item {[9]} D. Kong and W. Zhang, Phys. Lett. A {\bf 190}
(1994) 155.
\item {[10]} W. Zheng, J. Phys. A {\bf 27} (1994) L931.
\item{[11]} B. Dey, J. Phys. C: Solid State Phys. {\bf19} (1986) 3365;
E. Magyari, Z. Phys. B {\bf43} (1981) 345.
\item{[12]} A. Khare, Lett. Math. Phys. {\bf 3} (1979) 475.
\item{[13]} N. Akhmediev and V.V. Afansjev, Phys. Rev. Lett. {\bf 75} (1995)
2320.

\vfill
\eject
\end